\documentclass[pre,aps,amsmath,amssymb,amsfonts,floatfix,superscriptaddress,showpacs,footinbib,twocolumn,longbibliography]{revtex4-1}

\usepackage{graphicx}
\usepackage{amsmath,amsfonts}
\usepackage{xcolor}
\usepackage{color}
\usepackage{mathtools}
\usepackage{pgfplots}
\usepackage{comment}
\usepackage{soul}
\usepackage[normalem]{ulem}
\usepackage{bm}

\newcommand{\Kv}{\ensuremath{\mathbf{K}}}

\newcommand{\Qv}{\ensuremath{\mathbf{Q}}}

\usepackage{float}
\usepackage{tikz}
\usetikzlibrary{calc}
\usetikzlibrary{decorations.pathmorphing}
\usetikzlibrary{decorations.pathreplacing}
\usetikzlibrary{decorations.markings}
\usetikzlibrary{shapes.misc}
\usetikzlibrary{positioning}

\usetikzlibrary{arrows}
\usetikzlibrary{fadings}
\usetikzlibrary{shapes.callouts,tikzmark}
\tikzfading[name=fade out,
inner color=transparent!0,
outer color=transparent!100]

\tikzset{snake it/.style={decorate, decoration=snake}}
\usetikzlibrary{3d}
\usepackage{mathtools}

    \tikzset{
            partial ellipse/.style args={#1:#2:#3}{
                        insert path={+ (#1:#3) arc (#1:#2:#3)}
                            }
                        }

\usetikzlibrary{calc}
\tikzset{
            inertial frame/.style = {x={(-20:2cm)}, y={(-160:2cm)}, z={(90:2cm)}},
              local frame/.style = {shift={(local origin)}, x={(40:.7cm)}, y={(150:.7cm)}, z={(105:.7cm)}}
          }
\tikzset{middlearrow/.style={
                decoration={markings,
                            mark= at position 0.5 with {\arrow{#1}} ,
                                    },
                                            postaction={decorate}
                                                }
                                                }
\tikzset{cross/.style={cross out, draw, 
         minimum size=2*(#1-\pgflinewidth), 
                  inner sep=0pt, outer sep=0pt}}

\def\presuper#1#2%
  {\mathop{}%
   \mathopen{\vphantom{#2}}^{#1}%
   \kern-0.5\scriptspace%
   #2}

\usepackage{hyperref}
\hypersetup{colorlinks=true,breaklinks,linkcolor=blue,urlcolor=blue,citecolor=blue}

\begin{document}
\def\umphys{%
    Department of Physics, University of Michigan,
    Ann Arbor, Michigan 48109, USA
}%
\def\tuwien{
Institute for Solid State Physics, TU Wien, 1040 Vienna, Austria
}

\author{Yang Yu}
\affiliation{\umphys}
\author{Sergei Iskakov}
\affiliation{\umphys}
\author{Emanuel Gull}
\affiliation{\umphys}
\author{Karsten Held}
\affiliation{\tuwien}
\author{Friedrich Krien}
\affiliation{\tuwien}

\title{Pairing boost from enhanced spin-fermion coupling in the pseudogap regime}

\begin{abstract}
We perform a fluctuation analysis of the pairing interaction in the hole-doped Hubbard model within the dynamical cluster approximation.
Our analysis reveals that spin-fluctuation-mediated pairing differs qualitatively in the over- and underdoped regimes.
In the underdoped regime, spin fluctuations open a pseudogap.
We show that in this regime the spin-fermion coupling mediates a giant attraction between antinodal fermions. This explains why superconductivity survives at underdoping in the Hubbard model and cuprates, 
despite the lack of coherent fermionic quasiparticles in the pseudogap regime.
\end{abstract}

\maketitle
{\textit{Introduction.}---}
A fundamental question of condensed matter physics is why some of the highest superconducting transition temperatures are observed in doped Mott insulators~\cite{Imada98,Lee06,Anderson07, Keimer15}, where strong correlations destroy fermionic coherence and may open a pseudogap in the fermionic spectral function~\cite{Huscroft01,Maier02,Maier03,Parcollet04,Civelli05,Macridin06,Kyung06,Stanescu06,Haule07-2,Civelli08,Gull08-2, Werner09, gull09, Gull10, Kuchinskii12, Gull12, Tremblay12-2, Gull13,LeBlanc13,Gunnarsson15,Wu18,Schafer21,Ye23,Simkovic24}.
Indeed, within Migdal-Eliashberg theory the superconducting transition temperature depends exponentially on the pairing interaction $V$
and on the spectral weight at the Fermi level $N_0$: $T_c~\propto\exp[-2/(VN_0)]$.
Accordingly, the pseudogap is often considered to be an adversary to high-temperature superconductivity~\cite{Norman05,Norman14}.

A candidate for the pairing glue $V$ in cuprates are spin fluctuations~\cite{Kampf90,Pines93,Chubukov97,Abanov03,Prelovsek05,Scalapino12,Metzner19,Kitatani19,Restrepo22,Dong22-2}.
Theoretical approaches for the Hubbard model often perceive a competition between the pseudogap and superconductivity because the pairing glue itself, that is, the spin fluctuations, opens the pseudogap, thus undermining fermionic coherence~\cite{Abanov01}.
For example, based on a Migdal-Eliashberg-like ansatz using the spectral function of the experiment as the input, 
it was argued that pairing mediated by spin fluctuations is inefficient when superconductivity and the pseudogap have a common origin~\cite{Norman14}\footnote{As a possible way out, Ref.~\cite{Norman14} noted the possibility that the pseudogap is caused by superconductivity.}.
On the other hand, from Hubbard model calculations using the dynamical cluster approximation (DCA~\cite{Maier05,Hettler98,Hettler2000}), Maier and Scalapino concluded that the pairing through spin fluctuations remains strong despite the pseudogap~\cite{Scalapino16}.

The strength of the pairing interaction depends on the mediating boson, as well as on the fermion-boson coupling~\cite{Grilli91,Grilli94}.
A large coupling, therefore, may boost superconductivity.
Indeed, the highest critical temperatures are observed in conventional superconductors with large electron-phonon couplings~\cite{Drozdov15,Errea20}.
In the context of spin fluctuations, the quasistatic limit of the spin-fermion model suggests that
vertex corrections could enhance the spin-fermion coupling~\cite{Schmalian99};
however, a more widely accepted viewpoint is that they suppress it~\cite{Schrieffer95,Chubukov97,Huang05,Huang06,Maier07,Kitatani19}.

Here, we consider the pairing interaction in the hole-doped Hubbard model, often considered the quintessential model for cuprate superconductors~\cite {Anderson1987},
using the DCA
with auxiliary field~\cite{gull08} continuous-time quantum Monte Carlo~\cite{rubtsov05,werner06,gull11} and submatrix updates~\cite{gull11-2}.
For the precise tiling of the momentum space in DCA, see the Supplemental Material~\cite{suppl}\nocite{Scalapino1995,Schaefer13, Gunnarsson16,Schafer16, Vucicevic18, Krien22-2, Kitatani22,vanLoon20} and Ref.~\cite{Gull10}.
We compute explicitly the spin-fermion coupling in the under- and overdoped regimes and examine the role it plays for superconductivity.
Our results show that the increased scattering and decoherence of antinodal fermions in the pseudogap regime
is compensated by a concomitant increase of their coupling to spin fluctuations.
Indeed, without this large spin-fermion coupling,
$d$-wave pairing fluctuations would be strongly suppressed in this regime.
As a result, the pseudogap is, in fact, not an adversary to pairing,
but instead marks a regime of strongly correlated superconductivity where the spin-fluctuation pairing mechanism differs qualitatively from overdoping.

{\textit{Pairing glue and spin-fermion coupling.}---}
In a nutshell, our main result is that the spin-fermion pairing glue
\begin{eqnarray}
\frac{3}{2}U^2|\gamma_{KQ}|^2\chi_Q,\label{eq:glue}
\end{eqnarray}
with $Q=\pm K'-K$, depends crucially on the momentum-frequency $K=(\Kv,\nu)$ of the incoming fermion, the $K'$ of the outgoing one, and the bosonic $Q=(\Qv,\omega)$ of the spin fluctuations. Here, $\chi$ is the spin susceptibility, $U$ the Hubbard interaction, and $\gamma$ is the spin-fermion coupling vertex.
The latter is also known as the ``proper''~\cite{Giuliani05}, ``interaction-irreducible''~\cite{Hertz73}, or ``Hedin''~\cite{Hedin65} vertex (see Refs.~\cite{Krien19,Krien19-4} for a pedagogical introduction).

At low temperature, both $\gamma$ and $\chi$ peak at the antiferromagnetic wave vector $\mathbf{Q}=\mathbf{\Pi}=(\pi,\pi)$. More importantly, 
$|\gamma|$ is close to its non-interacting limit of $1$ only for nodal quasiparticles $[\Kv_N=(\pi/2,\pi/2)]$ or for overdoping,
whereas in the pseudogap regime the spin-fermion coupling 
becomes extremely large ($|\gamma|\gg 1$) for antinodal fermions $[\Kv_A=(\pi,0)]$ and a small Matsubara frequency $\nu$.
In fact, at low doping the glue between antinodal fermions may be enhanced by orders of magnitude due to $\gamma$.
As a consequence, pairing remains strong despite the pseudogap.

{\textit{Spin-fermion coupling.}---}
In our calculations we use the eight-site DCA~\cite{Maier05} to solve the Hubbard model with nearest-neighbor hopping $t\equiv 1$, next-nearest-neighbor hopping $t'=-0.15t$, and interaction $U=7t$ as employed before in Refs.~\cite{Werner09,gull09,lin10,Gull10,lin12,Dong19,Dong20,Yu24}.
For our doping scan, the temperature is fixed to $T=0.05t$, and we investigate hole dopings from $\delta=0$ to $0.25$.
In our temperature scan, we focus on $\delta=0.075$. From the DCA we obtain the cluster self-energy $\Sigma_K$, 
where $K=(\Kv,\nu)$ again denotes a fermionic momentum-frequency.
We also evaluate the spin-fermion vertex $\gamma_{KQ}$ and the susceptibility $\chi_Q$ of the cluster~\cite{Yu24}.
We focus on static antiferromagnetic fluctuations ($\omega=0$), which are dominant in the pseudogap regime~\cite{Gunnarsson16,Wu22}.

\begin{figure}
\centering
\includegraphics[width=0.47\textwidth]{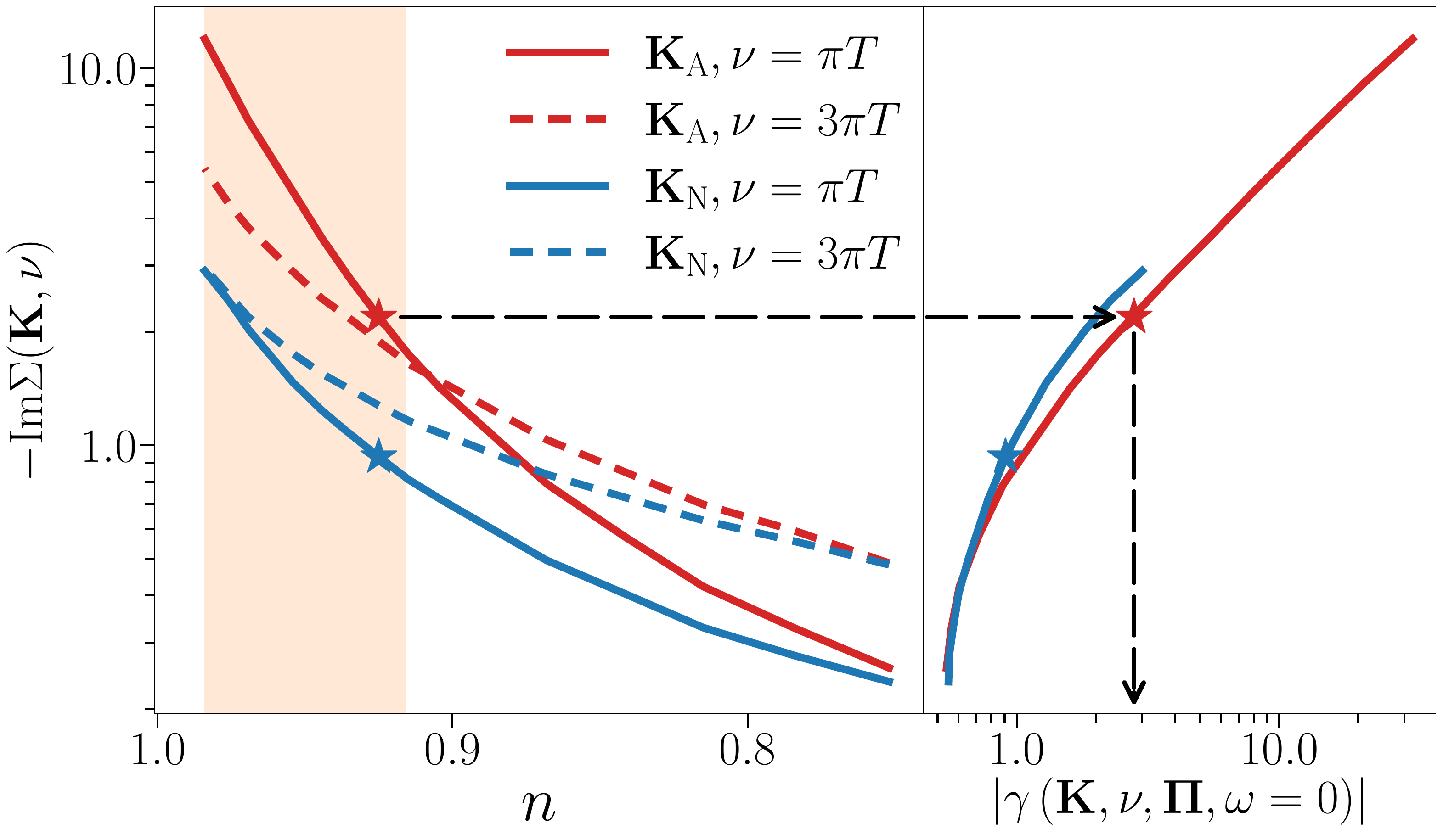}
\caption{Left: Imaginary part of the nodal (blue) and antinodal (red) self-energy 
at the first (solid) and second (dashed) Matsubara frequencies as a function of doping ($T=0.05t$), as calculated on an eight-site DCA cluster for the Hubbard model.
{The light orange background highlights the pseudogap regime.}
Right: Absolute value of the spin-fermion coupling at $\nu=\pi T$, $\omega=0$, and $\mathbf{Q}=\mathbf{\Pi}=(\pi,\pi)$ plotted against the imaginary part of the self-energy at the nodal (blue) and antinodal (red) points.
Stars mark a calculation at $\delta=0.075$, where $|\gamma|\sim3$ for antinodal fermions (cf. dashed arrows).}\label{fig:sigma_gamma} 
\end{figure}

The left panel of Fig.~\ref{fig:sigma_gamma} shows a  
doping scan of the self-energy at the first and second Matsubara frequencies from half-filling ($n=1$ electron per site) to $\delta = 0.25$ hole doping ($n=0.75$).
The antinodal self-energy (red line) shows insulating behavior, indicated by {$-\mathrm{Im}\Sigma(\Kv_A,\pi T)>-\mathrm{Im}\Sigma(\Kv_A,3\pi T)$ \cite{Schaefer15}},
below the doping $\delta=0.08$ {(light orange background)}, whereas the nodal self-energy (blue line) remains metallic in the doping range considered.

In the right panel of Fig.~\ref{fig:sigma_gamma}, we plot the absolute value of the coupling
$\gamma(\Kv,\nu=\pi T,\Qv=\bm{\Pi},\omega=0)$ of fermions to static antiferromagnetic fluctuations [$\omega=0$, $\bm{\Pi}=(\pi,\pi)$]
against $-\text{Im}\Sigma(\Kv,\nu=\pi T)$ ($y$-axis) at the corresponding fermionic momentum.
Strikingly, for large scattering they are connected by a power law.
For antinodal fermions, we find $|\gamma|\gg1$, orders of magnitude beyond the noninteracting limit $|\gamma|=1$, whereas for nodal fermions $|\gamma|\approx1$.

For example, stars in the left and right panels mark the calculation results at $\delta=0.075$ in the pseudogap regime.
The large self-energy at the antinode is accompanied by a coupling $|\gamma|\sim3$ (cf. dashed arrows) of antinodal fermions to spin fluctuations. Already for this doping, the pairing glue $V\sim|\gamma|^2$~\eqref{eq:glue} is enhanced by one order of magnitude.
In other words, the more correlated fermions are, the more strongly they couple to spin fluctuations,
leading to a nodal/antinodal dichotomy not only for the one-particle excitations in the form of the pseudogap, but, at the same time, also of the spin-fermion coupling.
Our results on a larger cluster at a higher temperature show that this dichotomy is likely underestimated on the eight-site cluster( see Supplemental Material~\cite{suppl}).
Notice that $\gamma$ has a sizable imaginary part~\cite{Krien22},
but only its absolute value enters the pairing glue~\eqref{eq:glue}~\cite{suppl}. 

{\textit{Superconductivity.}---}
The static pairing susceptibility $\hat{P}$ in the normal state can be calculated from the Bethe-Salpeter equation
$\hat{P}=-\hat{\chi}^0-\frac{1}{2}\hat{\chi}^0\hat{\Gamma}\hat{P}$. Here, $\chi^0_{KK'}$ is the coarse-grained bare bubble and $\Gamma_{KK'}$ the (particle-particle) irreducible vertex of the DCA cluster.
All quantities are matrices in the coarse-grained momentum-frequencies $K,K'$.
Often one considers the eigenvalue $\lambda$ of $1/2\;\hat{\chi}^0\hat{\Gamma}$ with
$\lambda\rightarrow 1$ signaling the superconducting phase transition \cite{Maier06,Maier06-2,Maier07,Maier07-2,Scalapino12}, and we do so in the Supplemental Material~\cite{suppl}.

Here, instead, we examine the contribution of spin fluctuations to the pairing susceptibility~\cite{Maier05-2,Gull15}
using a modified Bethe-Salpeter equation.
To this end, we separate the irreducible vertex into two parts, $\Gamma=\Gamma^{\mathrm{sf}}+\Gamma^{\mathrm{r}}$,
where $\Gamma^{\mathrm{sf}}$ captures the spin-fluctuation exchange and $\Gamma^{\mathrm{r}}$ the rest.
The former is given as~\cite{Krien20-2},
\begin{align}
\Gamma^{\mathrm{sf}}_{KK'}=&-\frac{3}{2}|\gamma_{K,-K'-K}|^2 W_{-K'-K}\notag\\&-\frac{3}{2}|\gamma_{K,K'-K}|^2 W_{K'-K}-3U,
\label{eq:gamma_sf}
\end{align}
where $W_Q=-U+U^2\chi_Q$ is the screened interaction.

We then calculate the pairing susceptibility in two steps~\cite{suppl}.
First, only the contribution of $\Gamma^{\mathrm{r}}$ is taken into account, and
this renormalizes the bare bubble $\chi^0$ to $\chi^r$.
{We observe in our calculations that $\Gamma^{\mathrm{r}}$ suppresses superconductivity in the  pseudogap regime.}
Second, we add the spin fluctuation exchange captured by $\Gamma^{\mathrm{sf}}$
and obtain the full pairing susceptibility as follows,
\begin{eqnarray}
\hat{P}=\left[\hat{1}-\frac{1}{2}\hat{\chi}^r\hat{\Gamma}^{\mathrm{sf}}\right]^{-1}\hat{\chi}^r.\label{eq:bse_mod}
\end{eqnarray}
Since $\hat{P}$ is solved via matrix inversion, this calculation can be performed reliably even when an iterative solution of the Bethe-Salpeter equation fails, which is often the case in strongly correlated regimes~\cite{suppl}.

For $d$-wave pairing, we split $\hat{P}$ into the  vertex and bare bubble contribution, and project onto the corresponding $d$-wave form factor:
$\mathcal{P}^d=-\frac{4T^2}{N_c^2}\sum_{KK'}g^d_{\Kv}({P}_{KK'}+\chi^0_{KK'})g^d_{\Kv'}$
and $P^d_0=\frac{4T^2}{N_c^2}\sum_{KK'}g^d_{\Kv}\chi^0_{KK'}g^d_{\Kv'}$.
Here, $N_c=8$ is the size of the cluster and $g^d_{\Kv}=\cos K_x - \cos K_y$ is the $d$-wave symmetry factor;
a minus sign aligns our definition of $\hat{P}$ with the physical pairing susceptibility.
Notice that $\mathcal{P}^d$ may be negative, and
in this case the vertex corrections describe a repulsion in the $d$-wave channel.

\begin{figure*}
\centering
\includegraphics[width=0.88\textwidth]{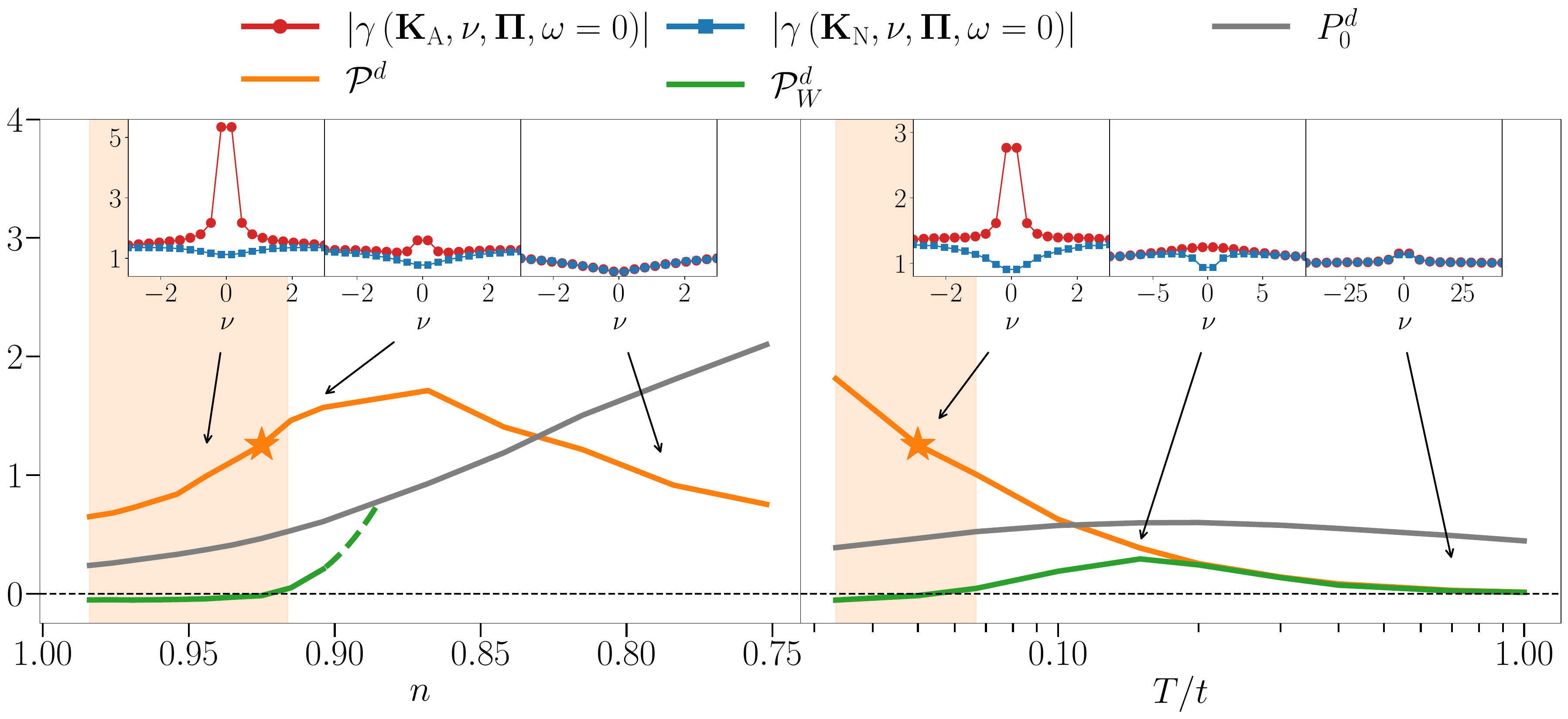}
\caption{Left: $d$-wave pairing susceptibility $\mathcal{P}^d$   (orange line) as a function of doping ($T=0.05t$).
The green line shows the susceptibility $\mathcal{P}^d_W$ for an artificial constant $|\gamma|\equiv1$,
which diverges for this temperature in the overdoped regime (the dashed green line shows the onset of this divergence).
A gray line shows the bare pairing susceptibility.
Right: The same quantities shown as functions of temperature ($\delta=0.075$).
Insets: Absolute value of the spin-fermion coupling for parameters indicated by arrows.
All results are shown for the Hubbard model in eight-site DCA with parameters $t=1$,  $t'=-0.15t$, and $U=7t$.
The light orange hue highlights the pseudogap regime;
stars mark a representative point $(\delta=0.075, T=0.05t)$ (cf. Fig.~\ref{fig:sigma_gamma}).
}\label{fig:chi}

\end{figure*}

Equations~\eqref{eq:gamma_sf} and~\eqref{eq:bse_mod} allow us to pinpoint the contribution of $\gamma$ to the pairing susceptibility.
Namely, if we leave $\gamma$ in Eq.~\eqref{eq:gamma_sf} as is, we obtain through Eq.~\eqref{eq:bse_mod}
the exact pairing susceptibility corresponding to the DCA.
The orange line in the left panel of Fig.~\ref{fig:chi} shows the corresponding $\mathcal{P}^d$ as a function of doping at $T=0.05t$.
Similar to previous investigations~\cite{Gull15} it is positive and has a dome shape.
The $d$-wave projection of the bare bubble, $P^d_0$,
is suppressed in the pseudogap regime (see the gray line).
Indeed, the opening of the pseudogap cuts off the divergence of the bubble at low temperature~\cite{Scalapino16}.

Let us now consider the relevance of the spin-fermion coupling $\gamma$.
The insets in Fig.~\ref{fig:chi} show $\gamma(\Kv,\nu,\Qv=\bm{\Pi},\omega=0)$ as a function of $\nu$ for antinodal (red) and nodal (blue) fermions.
For overdoping, $\gamma$ depends only weakly on the fermionic momentum $\Kv$ and is suppressed below $1$ at low frequencies.
Near optimal doping momentum dependence becomes appreciable and for antinodal fermions $|\gamma|$ is enhanced at low frequencies.
For underdoping, $\gamma$ exhibits a large nodal/antinodal dichotomy.
In this regime the pairing of antinodal fermions is greatly enhanced by $|\gamma|^2$,
explaining the still appreciable value of $\mathcal{P}^d$, despite the opening of a pseudogap.

To highlight the crucial importance of the spin-fermion coupling, we set $\gamma\equiv1$
in Eq.~\eqref{eq:gamma_sf} and calculate the pairing susceptibility once again via Eq.~\eqref{eq:bse_mod};
we refer to its $d$-wave projection (with again $\chi^0$ subtracted) as $\mathcal{P}^d_W$~\cite{suppl}.
The result (green line) shows that in the pseudogap regime $d$-wave fluctuations are virtually {\sl absent} without the enhancement of $|\gamma|$ for antinodal fermions.
Conversely, for larger dopings, $|\gamma|$ is suppressed below $1$ (see insets) and setting it to $1$ enhances the pairing glue.
Without this suppression of the spin-fermion coupling $\gamma$, $\mathcal{P}^d_W$ actually diverges for the chosen parameters (dashed).

The right panel of Fig.~\ref{fig:chi} shows the trends as a function of the temperature at $\delta=0.075$.
At high temperature, $\gamma\approx1$ and hence $\mathcal{P}^d\approx\mathcal{P}^d_W$.
We observe a small enhancement $|\gamma|\gtrsim1$ but no momentum dependence (right inset).
At low temperature, we observe the described strong enhancement of $\gamma$ at the antinode (left inset). 
$\mathcal{P}^d$ (orange) increases as the temperature decreases, whereas $\mathcal{P}^d_W$ (green) becomes vanishingly small.
Stars in the left and right panels denote the parameters $\delta=0.075, T=0.05t$, also highlighted with stars in Fig.~\ref{fig:sigma_gamma}.

{\textit{Interpretation.}---}
Our DCA calculations reveal two regimes of spin-fluctuation-mediated superconductivity.
In the overdoped regime, correlations are weak and $\gamma$ is suppressed.
In the underdoped regime, antinodal fermions are strongly damped by a large $\text{Im}\Sigma$, a hallmark of the pseudogap regime;
however, the same strong coupling to antiferromagnetic spin fluctuations that leads to this damping in the first place, also enhances the coupling of fermions in the Cooper channel.
The first (overdoped) case can be considered a weak correlation regime of superconductivity 
because $\gamma$ is generally suppressed at small $U$, also for small dopings~\cite{suppl}.
A suppression is also characteristic of the Fermi liquid regime, as a result of particle-particle screening~\cite{Krien20,Harkov21}.
In the following we explain why the second case may be referred to as a strong correlation regime of superconductivity. 

First, let us note that an antinodal enhancement of $\gamma$ is similar to the quasistatic limit of the spin-fermion model~\cite{Schmalian99}.
In that case $\gamma$ is enhanced $\propto\xi^2$, where $\xi$ is the correlation length. 
However, the latter remains small in the Hubbard model at strong coupling, even in the absence of frustration~\cite{Simkovic22}.
Instead, we argue that incipient localization is the driver of a large $\gamma$.
This is most easily seen in the case of the atomic limit~\cite{Krien19-3}, 
where $|\gamma|$ diverges in the limit $T\rightarrow0$ and is connected to $-\text{Im}\Sigma$ via a power law~\cite{suppl},
similar to the right panel of Fig.~\ref{fig:sigma_gamma}.
More generally, $\gamma$ is enhanced in the limit of an unscreened local moment,
as was previously observed~\cite{Katanin09,vanLoon18,Harkov21,Adler24} in the context of the dynamical mean-field theory (DMFT~\cite{Georges96}).
In fact, Ref.~\cite{Harkov21} showed that the enhancement of $\gamma$ mediates the effective exchange $4t^2/U$.

Moving on to nonlocal correlations, DCA is a natural extension of DMFT.
Here, the pseudogap emerges as antinodal fermions decouple from the DCA's effective medium,
eliminating Kondo screening in a momentum-dependent fashion~\cite{Gunnarsson14}.
Our results show that the decoupling from the DCA  bath also leads to the enhancement of $\gamma$ in the antinodal direction.
All these effects hinge on a large Hubbard interaction and incipient localization near half filling.
Therefore, the enhancement of the pairing glue by a factor $|\gamma|^2\gg1$ is a peculiar strong correlation phenomenon.

With regard to the power law observed when drawing $|\gamma|$ against $-\text{Im}\Sigma$,
we note that they are connected via Hedin's equation $\Sigma\sim GW\gamma$, where $G$ is the Green's function.
Upon a breakdown of Kondo screening, the emergence of poles in the self-energy
is ubiquitous in a variety of models of strongly correlated electrons~\cite{Pavarini17}. 
Indeed, such a pole is needed to open a (pseudo)gap without symmetry breaking or order.
We observe here that, for the underdoped Hubbard model in two dimensions, $\Sigma$ and $\gamma$ could form poles concomitantly; this raises the intriguing possibility that pairing may be sustained in the absence of a Fermi surface, namely, by a Luttinger surface~\cite{Fabrizio22,Fabrizio23,Worm23}.

{\textit{Conclusions.}---}
We computed the spin-fermion coupling $\gamma$ within the eight-site dynamical cluster approximation (DCA) for the hole-doped Hubbard model in the pseudogap regime. We find that $\gamma$ exhibits a significant nodal/antinodal dichotomy in momentum space. The pairing glue is thus enhanced by a large factor $|\gamma|^{2}$ for antinodal fermions, even though these fermions lose their quasiparticle character due to the opening of the pseudogap. Without this enhancement, $d$-wave pairing would not occur in the underdoped regime. 

Previous studies of the spin-fluctuation glue in the normal state often assumed an isotropic spin-fermion coupling $\gamma\approx1$ (e.g., Refs.~\cite{Maier07,Maier07-2,Norman14,Restrepo22}), which we find may only apply in weakly correlated systems or at high temperatures. 
With regard to the superconducting state, a recent study found that spin fluctuations contribute only about half of the superconductivity in the strongly correlated regime, based on an isotropic $\gamma$~\cite{Dong22}.
It is plausible that for underdoping the dichotomy and antinodal enhancement of $\gamma$ observed here persist in the superconducting state~\footnote{Note that spin fluctuations may change qualitatively in the superconducting state~\cite{Inosov09}.}, calling for further investigation of the lightly doped Hubbard model.
\newline

We thank Alessandro Toschi and Matthias Reitner for useful comments. Y.Y. and E.G. were supported by the National Science Foundation under Grant No. NSF DMR 2401159; F.K. and K.H. by the Austrian Science Fund (FWF) projects  P32044,   P36213, V1018, SFB Q-M\&S (FWF project ID F86), and Research Unit QUAST by the Deutsche Foschungsgemeinschaft (DFG; project ID FOR5249) and FWF (project ID I 5868).

\bibliography{main}
\end{document}